\documentclass[iop,apj]{emulateapj}

\usepackage{amsmath,amstext}
\usepackage{graphicx}
\usepackage{booktabs}

\usepackage[breaklinks,colorlinks,citecolor=blue,linkcolor=magenta]{hyperref} 
\usepackage[all]{hypcap} 

\begin{document}

\title{Constraining the Outflow Structure of the Binary Neutron Star Merger Event GW170817 with Markov-Chain Monte Carlo Analysis}
\author{Yiyang Wu and Andrew MacFadyen}
\affil{Center for Cosmology and Particle Physics, New York University}

\begin{abstract}
The multi-wavelength non-thermal emission from the binary neutron star (BNS) merger GW170817 has raised a heated debate concerning the post-merger outflow structure. Both a relativistic structured jet viewed off-axis and a mildly relativistic quasi-spherical outflow can explain the observational data of GW170817 up to $\sim260$ days. We  utilize a physically motivated analytic two-parameter model called the ``boosted fireball'', for the outflow structure after it has expanded far from the merger site and has entered the self-similar coasting phase. This model consists of a family of outflows with a structure varying smoothly between a highly collimated ultra-relativistic jet and an isotropic fireball. We simulate the dynamical evolution, starting with ``boosted fireball'' initial conditions, of $240$ outflows using the moving-mesh relativistic hydrodynamics code \texttt{JET} to follow their evolution through the afterglow phase. We compute nearly $2,000,000$ synchrotron spectra from the hydrodynamic simulations using the standard synchrotron radiation model. By making use of scaling relations in the hydrodynamic and radiation equations, we develop a synthetic light curve generator with an efficient sampling speed. This allows us to fit the observational data by performing Markov-Chain Monte Carlo (MCMC) analysis in a 8-dimensional parameter space, consisting of hydrodynamic, radiation and observational parameters. Our results favor the relativistic structured jet, with a jet opening angle  $\theta_0 \approx 5$ deg and Lorentz factor  $\Gamma \approx 175$, viewed from an off-axis angle $\theta_{obs} = 27^{+9}_{-3}$ deg. Due to parameter degeneracies, we find broad distributions for the explosion energy $E_0$, the circumburst density $n_0$, the electron energy fraction $\epsilon_e$ and the magnetic energy fraction $\epsilon_B$. The combination of a high $n_0$ and a low $\epsilon_B$ can also produce a good fit, indicating that very low $n_0$ may not be required for GW170817. 

\end{abstract}
\keywords{gravitational waves - gamma-ray bursts - hydrodynamics}

\maketitle
\section{Introduction}
The joint discover of gravitational waves \citep{Abbott2017} and multi-wavelength electromagnetic emission from the binary neutron star (BNS) merger GW170817 opened up a new era of the multi-messenger astrophysics.  The observations of a pulse of gamma-rays \citep{Goldstein2017, Savchenko2017} and a delayed non-thermal emission component, ranging from radio to X-rays \citep{Alexander2017, Haggard2017, Hallinan2017, Kasliwal2017, Margutti2017, Troja2017, ruan2018brightening, Margutti2018, Alexander2018, Dobie2018, nynka2018fading}, provide convincing evidence  that BNS mergers are associated with short gamma-ray bursts (sGRBs), though of low detected fluence in the case of GRB170817A. The non-thermal emission following GW170817 differs from classical GRB afterglows \citep{Nakar2018}. Early observations resulted in non-detections \citep{Alexander2017}, until a first X-ray detection at $\sim 9$ days \citep{Troja2017} and a first radio detection at $\sim 16$ days \citep{Hallinan2017}. Once detected, the light curves exhibited a power-law brightening up to  $\sim 100$ days post-merger \citep{Hallinan2017, mooley2018mildly, ruan2018brightening}. Recent observations ($\sim 260$ days post-merger) indicated a turnover at $\sim 150$ days\citep{Alexander2018, Dobie2018, nynka2018fading}. 

In contrast to GRB170817A, most GRB afterglows are detected shortly after the prompt burst and peak at a very early time.  The GRB afterglow picture has been reviewed extensively \citep[e.g.,][]{zhang2004gamma, piran2005physics, 2018arXiv180101848V}. In the standard fireball model \citep{Goodman1986,Paczynski1986} a hot explosion expands and compresses itself into a thin relativistic shell. After most of its thermal energy is transfered to kinetic energy, the shell coasts for a while and maintains a self-similar structure. This typically occurs before an amount of mass sufficient to decelerate the shell has been swept up. Eventually the shell decelerates driving an external shock into the circumburst medium which gives rise to the afterglow emission via synchrotron radiation. 
A simple model incorporating asymmetry of the outflow is the top-hat jet model. It is an angular truncation of a spherically-symmetric self-similar Blandford-McKee blast wave solution \citep{blandford1976fluid}. The top-hat jet model is characterized by a uniform angular distribution within a jet opening angle $\theta_0$ with sharp edges at the boundaries. By using the top-hat jet model as the initial condition for hydrodynamical simulations, \citet{Eertenn2012} developed a synthetic light curve generator, the \texttt{BoxFit} package, which provides a useful tool to study GRB afterglows. However, the top-hat jet model is an oversimplification for  the outflow structure of GW170817. It cannot naturally account for the mild and steady rise of the non-thermal emission \citep{Kasliwal2017, Troja2017,mooley2018mildly, Margutti2018}. 

More complex models have been proposed to explain the properties of GW170817, which can be roughly classified into two categories. The first category of models, characterized by a mildly relativistic quasi-spherical outflow, includes the jet-less fireball model \citep{Salafia2018}, the choked jet-cocoon model \citep{Kasliwal2017,Bromberg2017, Gottlieb2018, mooley2018mildly, nakar2018gamma} and the fast component of the dynamical ejecta from the merger \citep{Hotokezaka2018}. The second category of models, characterized by a relativistic narrow core surrounded by slower wide-angle wings viewed from a substantial off-axis angle, consists of  the structured jet model \citep{Kathirgamaraju2017, Lamb2017, lazzati2017, d2018evidence, Margutti2018, Xie2018}, the Gaussian shaped jet model \citep{Troja2018,Resmi2018,Gill2018} and the successful jet-cocoon model \citep{Kasliwal2017}. These two categories of models can both explain the spectral and temporal evolution of the non-thermal emission. However,  they cannot be easily distinguished \citep{Dobie2018, Alexander2018, nynka2018fading}. Due to the complexity and variation of these models, it is challenging to systematically compare the two kinds of outflows or search for other outflow structures which may fit the data equally well or better. Some models are not physically motivated, for example assuming a Gaussian or a power low profile. Some models impose the jet opening angle by an arbitrary truncation. Thus, there is a lack of a simple and physically motivated model for generic outflows from GW170817. 

In this work, we consider an analytic, physically motivated, two parameter ``boosted fireball model,'' \citep{Duffell2013} to specify the outflow structure after it has reached the coasting phase far from the merger site. Instead of truncating a standard fireball into a top-hat, we launch a fireball with specific internal energy $\eta_0 \sim E/M$ with a bulk Lorentz factor $\gamma_B$ to describe an outflow characterized by energy and momentum injection that has occurred on the scale of the merger but which has subsequently expanded by many orders of magnitude in radius. Parameterized by these two physical quantities, the boosted fireball model is able to generate a family of outflows, with an isotropic fireball at one extreme and a highly collimated ultra-relativistic outflow at the other. The angular structure of the outflow is determined only by the two physical quantities. The jet opening angle arises naturally rather than in an ad hoc way.  

Similar to the top-hat jet model, the two-parameter boosted fireball model describes the outflow in the self-similar phase, when the outflow is already far away from the central engine. Due to the self-similarity,  we can apply scaling invariance in the hydrodynamic equations, which greatly reduces the simulation time. The simplicity of the two-parameter boosted fireball model allows us to develop a synthetic light curve generator and a Markov-Chain Monte Carlo (MCMC) sampler to fully explore the parameter space, similar to procedure used by \citet{Ryan2015} for the top-hat Blandford-McKee model. Compared to \citet{Ryan2015}, we replace the jet opening angle parameter from the top-hat jet model with the two parameters, $\eta_0$ and $\gamma_B$, from the boosted fireball model.  Instead of arbitrarily fixing some parameters and manually varying other parameters, our tools can fully explore the whole parameter space and automatically find the best-fitting parameters and their distributions. The best-fitting values for  $\eta_0$ and $\gamma_B$  characterize the outflow structure. The two categories of outflows can thus be naturally distinguished and other possibilities can be explored. 

In Section \ref{sec:bf}, we describe the basic properties of the two-parameter boosted fireball model and the setups for the relativistic hydrodynamic simulations. Section \ref{sec:fit} describes the synthetic light curve generation and MCMC curve fitting procedure. The results are summarized in Section \ref{sec:result} and discussed in Section \ref{sec:discussion}.


\section{Two-Parameter Boosted Fireball Model}
\label{sec:bf}
We first summarize the basic properties of the two-parameter boosted fireball model in Section \ref{sec: boosted}. In Section \ref{sec: hydro}, we describe the hydrodynamic simulation setups.

\subsection{Two-Parameter Boosted Fireball Model}
\label{sec: boosted}
The two-parameter boosted fireball model is a simple and physically motivated analytic model representing a family of outflows after they have expanded many orders of magnitude larger than the scale of the central engine, but before a significant amount of mass has been swept up. A fireball of specific internal energy $E/M$ is launched with a bulk Lorentz factor $\gamma_B$. In the center-of-momentum frame of the fireball, it expands isotropically and compresses itself into a thin shell with the asymptotic Lorentz factor $\eta_0 \sim E/M$. In the lab frame, the fireball is beamed in the launch direction with on-axis characteristic Lorentz factor $\Gamma \sim 2\eta_0\gamma_B$ and  characteristic jet opening angle $\theta_0 \sim 1/\gamma_B$. The jet opening angle arises naturally, which automatically satisfies the causality constraint $\Gamma  \gtrsim 1/\theta_0$. 

\begin{figure*}[hbt]
\label{fig: Rho}
\centering
\includegraphics[width=\linewidth]{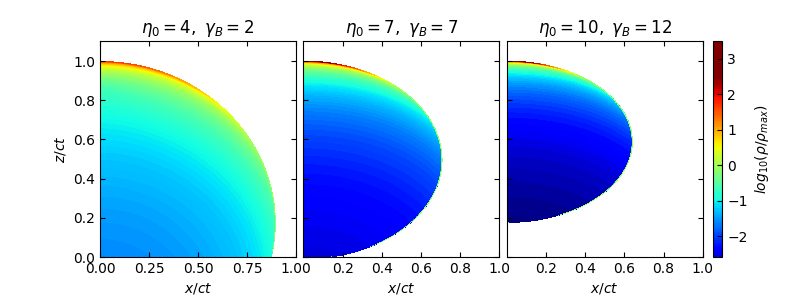}
\caption{Density profiles on a logarithmic scale of boosted fireballs for $\eta_0=4$ ,  $\gamma_B=2$  (left), $\eta_0=7$, $\gamma_B=7$ (middle) and  $\eta_0=10$ ,  $\gamma_B=12$ (right). The fireballs are boosted along the $z$ direction and have expanded far away from the central engine ($\sim 10^{15} \text{cm}$). The coordinates $x$ and $z$ are scaled by $ct$ due to self-similarity, where t is typically $10^{5} \text{s}$.  $\rho_{max}$ is a normalization constant. For different values of $\eta_0$ and $\gamma_B$, the boosted fireball model can generate different outflow structures ranging from a mildly relativistic quasi-spherical outflow with $\Gamma \sim 16$ (left) to an ultra-relativistic structured jet with $\Gamma \sim 240$ (right).} 
\end{figure*}


In Figure \ref{fig: Rho}, we show the density profiles for three typical boosted fireballs. By varying the values of $\eta_0$ and $\gamma_B$, all kinds of outflows with different jet opening angles and Lorentz factors can be generated. For  $\eta_0=4$ and  $\gamma_B=2$, the boosted fireball has a mildly relativistic quasi-spherical structure. In the choked jet-cocoon scenario, a jet is choked by the BNS ejecta and deposits its energy in the ejecta. The resulting system is less energetic and mildly relativistic, which can be reasonably approximated by a cooler fireball launched with a small Lorentz factor.
The boosted fireball of $\eta_0=7$ and  $\gamma_B=7$ is a collimated and relativistic outflow with  $\Gamma \sim 100$. A hot fireball launched with a large Lorentz factor is associated with the successful jet-cocoon scenario: a powerful jet successfully breaks out from the BNS ejecta and propagates with relativistic velocity. In  $\eta_0=10$ and  $\gamma_B=12$ case, the boosted fireball model can generate a collimated and ultra-relativistic structured jet with $\Gamma \sim 240$ . 

Due to relativistic beaming, the boosted fireball has a jet opening angle $\theta_0 \sim 1/\gamma_B$ and an explosion energy $E_{0} \sim \gamma_B\eta_0E$ (one side). The isotropic equivalent energy is estimated to be 
\begin{equation}\label{eq: Eiso}
E_{iso} \sim \frac{2E_0}{1-cos(\theta_0/2)}.
\end{equation}

We can derive the maximum Lorentz factor and the energy per solid angle  with respect to angle $\theta$ as follows (details see \citet{Duffell2013}): 
\begin{equation}\label{eq:Theta}
\gamma_{max}(\theta) =  \frac{\gamma_B\left(\eta_0+v_Bcos\theta\sqrt{\eta_0^2v_0^2 - \gamma_B^2 v_B^2 sin^2\theta}\right)}{1+\gamma_B^2v_B^2sin^2\theta}
\end{equation}
\begin{equation}
\frac{dE_0}{d\Omega} = \frac{E_{iso}}{4\pi} \left( \frac{\gamma_{max}(\theta)}{\gamma_{max}(0)} \right)^3, 
\end{equation}
where $v_0=\sqrt{1-1/\eta_0^2}$ and $v_B=\sqrt{1-1/\gamma_B^2}$. 

\begin{figure}[hbt]
\label{fig: TauU0}
\centering
\includegraphics[width=\linewidth]{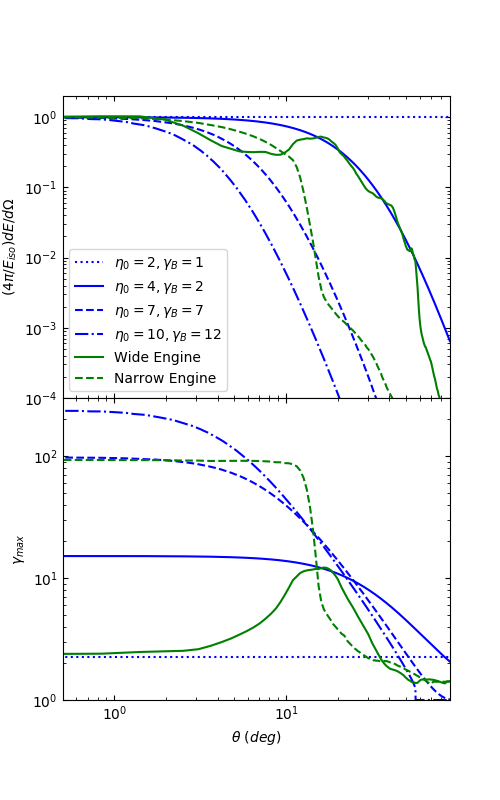}
\caption{We show the energy per solid angle (top) and the maximum Lorentz factor (bottom) as a function of angle $\theta$ for the isotropic fireball and the three typical boosted fireballs (blue lines). The green lines shows the profiles for the wide and narrow engine models from \citet{Xie2018}. The energy per solid angle is normalized for comparison. }
\end{figure}

We show the energy per solid angle and the maximum Lorentz factor as a function of $\theta$ for the isotropic fireball and the three typical boosted fireballs in Figure \ref{fig: TauU0}. For comparison, we also present the profiles for wide engine and narrow engine models from \citet{Xie2018} . \citet{Xie2018} simulated the complete life-cycle of the outflow, from the accelerating phase, through breakout and the coasting phase, to the decelerating phase. The boosted fireball model starts from the coasting phase, where the outflow has a self-similar structure. Thus, the profiles from \citet{Xie2018} provide insight to the boosted fireball model. As we can see in Figure \ref{fig: TauU0}, the boosted fireball model gives reasonable approximations to the wide engine and narrow engine models. Thus, the boosted fireball model can reasonably represent the mildly relativistic quasi-spherical outflow and the relativistic structured jet. 

The two-parameter boosted fireball model is a generic model, which can generate a family of outflows parameterized by $\eta_0$ and $\gamma_B$. In the limit $\gamma_B \rightarrow 1$, it becomes a standard spherical fireball. In the limit $\gamma_B \rightarrow \infty$, the outflow is an ultra-relativistic flow with a negligible jet opening angle. In between, it can generate a smoothly connected family of outflows with different Lorentz factors and jet opening angles. The generality of the boosted fireball model provides a systematic way to study the outflow structure and compare different kinds of outflows. In this study, we consider a large parameter space for the boosted fireball model with the characteristic Lorentz factor ranging from $2$ to $240$ and the characteristic jet opening angle ranging from $180^{\circ}$ to $5^{\circ}$. 

\subsection{Hydrodynamic Simulations}
\label{sec: hydro}
To study the evolution of boosted fireballs into the afterglow stage and compute their radiative signatures,  we perform two-dimensional relativistic hydrodynamic simulations using the moving-mesh code \texttt{JET} \citep{Duffell2013RT}. The radial motion of grid cells allows us to capture flow structures with high resolution and evolve flows with large Lorentz factors. We also implement adaptive mesh refinement to resolve detailed flow features, especially the thin shells and shear layers that develop in relativistic flows \citep{2006ApJS..164..255Z}.

In this study, we run a total of $240$ simulations with $\eta_0$ ranging from $2$ to $10$ with an increment of $1$ and $\gamma_B$  ranging from $1$ to $12$ with an increment of $0.5$. We employ a smaller increment for $\gamma_B$, since we find the synthetic light curves are more sensitive to $\gamma_B$.  We also insert extra points, $\eta_0=2.5$ and $\gamma_B = 1.25$, to ensure accuracy in the low $\eta_0$ and low $\gamma_B$ region.

Our initial condition is a boosted fireball of explosion energy $E_0$ expanding in a circumburst medium of constant number density $n_0$. The simulation begins long before a significant amount of mass has been swept up, at radius $r_{0}=0.01l$, where $l \equiv (E_{0}/m_p n_0)^{1/3}$ is the Sedov length with $m_p$ the proton mass. We check different values of the starting radius and find it does not significantly affect the dynamics as long as it is much smaller than Sedov length. We employ an equation of state with the adiabatic index changing smoothly from $4/3$ for a relativistic fluid to $5/3$ for a non-relativistic fluid \citep{Ryu2006}. 

Because the profiles for the boosted fireball are self-similar, we can take advantage of the scale invariance between $E_0$ and $n_0$ . We initially set  $E_{0}=10^{50} \text{erg}$ and $n_0=1  \text{ proton cm}^{-3}$,  corresponding to the starting radius $r_{0} \sim 10^{15}$ cm.  However, we can always rescale $E_0$ and $n_0$ to other values $E'_{0}$ and $n'_0$ using following relations: 
\begin{eqnarray}
E'_{0} &=& \kappa E_{0}\nonumber \\
n'_0  &=& \lambda n_0 \nonumber \\
r' &=&  (\kappa/\lambda)^{1/3}r \nonumber \\
t' &=& (\kappa/\lambda)^{1/3}t.
\end{eqnarray}
A detail description of scale invariance in hydrodynamic equations can be found in \citet{Eertenn2012Scale, Eertenn2012}. State-of-art relativistic hydrodynamic simulations of long time evolutions typically take days to run. Scale invariance greatly reduces the computing time, since only one simulation for each ($\eta_0$, $\gamma_B$) pair is needed and can be scaled for arbitrary values of $E_0$ and $n_0$.  The hydrodynamics of the outflow is fully parameterized by four parameters: the explosion energy $E_{0}$, the circumburst density $n_0$, the asymptotic Lorentz factor $\eta_0$ and the boost Lorentz factor $\gamma_B$.

\section{Fitting}
\label{sec:fit}
\subsection{Light Curve Generation}
To perform MCMC curve fitting of the observational data, we need to calculate millions of synthetic light curves with distinct hydrodynamic parameters, radiation parameters and observational parameters. For the two-parameter boosted fireball model , the hydrodynamic parameters consist of the explosion energy $E_{0}$, the circumburst density $n_0$, the asymptotic Lorentz factor $\eta_0$ and the boost Lorentz factor $\gamma_B$. For a standard synchrotron radiation model \citep{Piran1998}, the radiation parameters include the spectral index $p$, the electron energy fraction $\epsilon_e$, the magnetic energy fraction $\epsilon_B$ and the fraction of electrons accelerated by the shock $\xi_N$. The observational parameters are the redshift $z$, the luminosity distance $d_L$ and the observation angle $\theta_{obs}$. The high dimensionality makes it challenging to compute sufficient numbers of light curves on a computationally efficient timescale. 

To deal with the high-dimensionality, we adapt similar techniques used by \citet{Eertenn2012} and \citet{Ryan2015}. First, we compress the \texttt{JET} simulation snapshots  into so-called ``Box'' snapshots, which adequately capture all aspects of outflows but occupy a small amount of memory. The detailed procedure can be found in \citet{Eertenn2012}. The compression of snapshots allows us to load simulations with well sampled ranges of $\eta_0$ and $\gamma_B$ into memory at the same time. We implement bilinear interpolations between $\eta_0$ and $\gamma_B$, which allows us to quickly calculate light curves for arbitrary values of  $\eta_0$ and $\gamma_B$ within ranges specified in Table \ref{tab: table}. 

Second, we make use of scaling relations in the radiation equations. According to the standard afterglow model, the spectrum of synchrotron emission is a series of connected power laws, parameterized by the peak flux $F_{peak}$, the cooling frequency $\nu_c$ and the synchrotron frequency $\nu_m$ \citep{Piran1998}. Since synchrotron self-absorption does not play an important role in the current observations of GW170817, the synchrotron self-absorption frequency is omitted for simplicity. The scaling relations for observer time $t_{obs}$,  the peak flux $F_{peak}$, the synchrotron frequency $\nu_m$  and the cooling frequency $\nu_c$ are given by:
\begin{eqnarray}\label{eq: scale}
t_{obs} &=& (1+z)\left( \frac{E_{0}}{n_0} \right)^{1/3} \tau \nonumber \\
F_{peak} &=& \frac{1+z}{d_L^2}\frac{p-1}{3p-1}E_{0}n_0^{1/2}\epsilon_B^{1/2}\xi_N f_{peak}(\tau; \eta_0, \gamma_B, \theta_{obs})\nonumber  \\
\nu_m &=& \frac{1}{1+z} \left( \frac{p-2}{p-1} \right)^2 n_0^{1/2} \epsilon_e^2 \epsilon_B^{1/2} \xi_N^{-2} f_m(\tau; \eta_0, \gamma_B, \theta_{obs}) \nonumber \\
\nu_c &=& \frac{1}{1+z} E_{0}^{-2/3} n_0^{-5/6} \epsilon_B^{-3/2} f_{c}(\tau; \eta_0, \gamma_B, \theta_{obs}) , 
\end{eqnarray}
where $\tau$ is the scaled time and $f_{peak}$, $f_m$ and $f_c$ are the characteristic spectral functions \citep{Ryan2015}. The characteristic spectral functions  $f_{peak}$, $f_m$ and $f_c$ contain the spectral dependence of $\tau$, $\eta_0$, $\gamma_B$ and $\theta_{obs}$, which cannot be described by any analytic expressions. Other dependences, such as $E_{0}$ and $n_0$, can be scaled out according to Equation \ref{eq: scale}. Given the characteristic spectral functions, we can use the scaling relations to calculate the spectral parameters $F_{peak}$, $\nu_m$ and $\nu_c$, which fully describe the spectral model of synchrotron emission. As a result, we reduce the original high-dimensional problem to a lower dimension problem, which is to determine the $4$-dimensional characteristic spectral functions  $f_{peak}$, $f_m$ and $f_c$. 

We calculate nearly $2,000,000$ spectra using the radiation module of \texttt{BoxFit} package \citep{Eertenn2012}. For each spectrum with specific $t_{obs}$, $\eta_0$, $\gamma_B$ and $\theta_{obs}$, we can calculate the values of the scaled time $\tau$ and the spectral parameters $F_{peak}$, $\nu_m$ and $\nu_c$. Using scaling relations indicated in Equation \ref{eq: scale}, we determine the  characteristic spectral functions  $f_{peak}$, $f_m$ and $f_c$ as a function of $\tau$, $\eta_0$, $\gamma_B$ and $\theta_{obs}$ by tabulation. Three $50\times33\times45\times21$ tables for $f_{peak}$, $f_m$ and $f_c$ are created. The details of the tables are shown in Table \ref{tab: table}. 

With the characteristic spectral function tables, we can generate arbitrary light curves  in milliseconds. Given a set of parameters $\{E_{0}, n_0, \eta_0, \gamma_B; p, \epsilon_e, \epsilon_B, \xi_N; z, d_L, \theta_{obs} \}$, we find the time series of   $f_{peak}$, $f_m$ and $f_c$ in the characteristic function tables by interpolation. Then, we use the scaling relations (Equation \ref{eq: scale}) to calculate the time series of the corresponding spectral parameters $F_{peak}$, $\nu_m$ and $\nu_c$. Thus, the time evolution of the spectrum is determined and multi-band light curves can be rapidly produced. 

\begin{table}[hbt]\label{tab: table}
\centering
\caption{Parameter Ranges for Characteristic Spectral Functions Tables.}
\begin{tabular}{@{}lll@{}}
\toprule
Parameter     & Range                                       & Number \\ \midrule
$\tau$        & {[}$8.64\times10^{2},\ 8.64\times10^{7}${]} & $50$     \\
$\eta_0$      & {[}$2,\ 10${]}                               & $33$     \\
$\gamma_B$    & {[}$1,\ 12${]}                               & $45$     \\ 
$\theta_{obs}$ & {[}$0,\ 1${]}                             & $21$  \\  \bottomrule
\end{tabular}
\end{table}

\subsection{MCMC Analysis}
Using the MCMC method, we fit the observational data of GW170817 to the synthetic light curves generated from the characteristic spectral function tables. The observational data is taken from \citet{Alexander2018, Margutti2018}, containing radio, optical and X-ray up to $\sim260$ days after merger. To enhance the performance of MCMC fitting, transformations are performed on certain parameters. $E_0$ and $n_0$ are made dimensionless: $E_{0,50}\equiv E_0/10^{50}\text{erg}$ and $n_{0,0}\equiv n_0/1\text{ proton cm}^{-3}$. $E_{0,50}$, $n_{0,0}$, $\epsilon_e$ and $\epsilon_B$ are measured logarithmically. To reduce the dimensionality, the redshift and luminosity distance are set as $z=0.0973$ and $d_L = 39.5 \text{ Mpc}$ as listed in the NASA Extragalactic Database. Due to the well known degeneracy between $\epsilon_e$, $\epsilon_B$ and $\xi_N$, we fix  $\xi_N=1$. In the end, we have a set of transformed parameters for the MCMC fitting procedure, which we refers to collectively as $\Theta$:
\begin{equation}
\Theta = \{log_{10}E_{0,50}, log_{10}n_{0,0},  \eta_0, \gamma_B, \theta_{obs}, log_{10}\epsilon_e, log_{10}\epsilon_B, p \}.
\end{equation}

We set the prior distribution for each parameter in $\Theta$, except for $\theta_{obs}$ , as a uniform distribution within specified bounds.  The observation angle $\theta_{obs}$ is given a prior distribution proportional to $ \sin\theta_{obs}$, which accounts for the geometrical effect of the solid angle. The bounds of $\Theta$ are chosen to contain the phenomenologically interesting regions and be within the ranges of characteristic spectral  function tables.  The bounds of prior distributions can be found in Table \ref{tab: parameter}. 
\begin{table}[hbt]\label{tab: parameter}
\centering
\caption{Bounds for Prior Distributions.}
\begin{tabular}{@{}ll@{}}
\toprule
Parameter     & Range                           \\             \midrule
$log_{10}E_{0,50}$  &  [-6, 3] \\
$log_{10}n_{0,0}$       & [-6, 3]   \\
$\eta_0$      &  [2, 10]                     \\
$\gamma_B$    & [1, 12]                               \\ 
$\theta_{obs}$ & [0, 1]                         \\
$\log_{10}\epsilon_e$ & [-6, 0]  \\
$\log_{10}\epsilon_B$ & [-6, 0]  \\
$p$ &  [2, 2.5]  \\  \bottomrule
\end{tabular}
\end{table}

The posterior distribution of $\Theta$ is generated using the parallel-tempered affine-invariant ensemble sampler implemented in the \texttt{emcee} package \citep{Goodman2010,Foreman2013}. The sampler uses an ensemble of walkers moving simultaneously in the parameter space and is designed to maintain affine-invariance. Parallel tempering is  used to better sample the multi-modal distribution. This is important for our study, because we know both the quasi-spherical outflow and the relativistic structured jet both are promising models for GW170817. 

We set $10$ temperature levels and $100$ walkers per level for the sampler. The walkers are initialized in a small ball near the maximum of the posterior,  calculated through trial runs. We choose a burn-in of 10,000 iterations. The convergence of the MCMC chains is tested for the Gelman-Rubin statistic. Sampling is performed for 10,000 iterations.

\section{Results}
\label{sec:result}
\subsection{Full Parameter Space}
\label{subsec: full}
The two-parameter boosted fireball model can encompass the two kinds of outflows,  the relativistic structured jet and the mildly relativistic quasi-spherical outflow, as well as all intermediate outflows between a highly collimated ultra-relativistic jet and an isotropic fireball . By performing MCMC fitting, walkers fully explore the 8-dimensional parameter space and concentrate in the region corresponding to the best-fitting parameters. The MCMC analysis naturally discovers the best-fitting outflow structure and observation angle.

\begin{figure*}[hbt]
\label{fig: distribution_off}
\centering
\includegraphics[width=\linewidth]{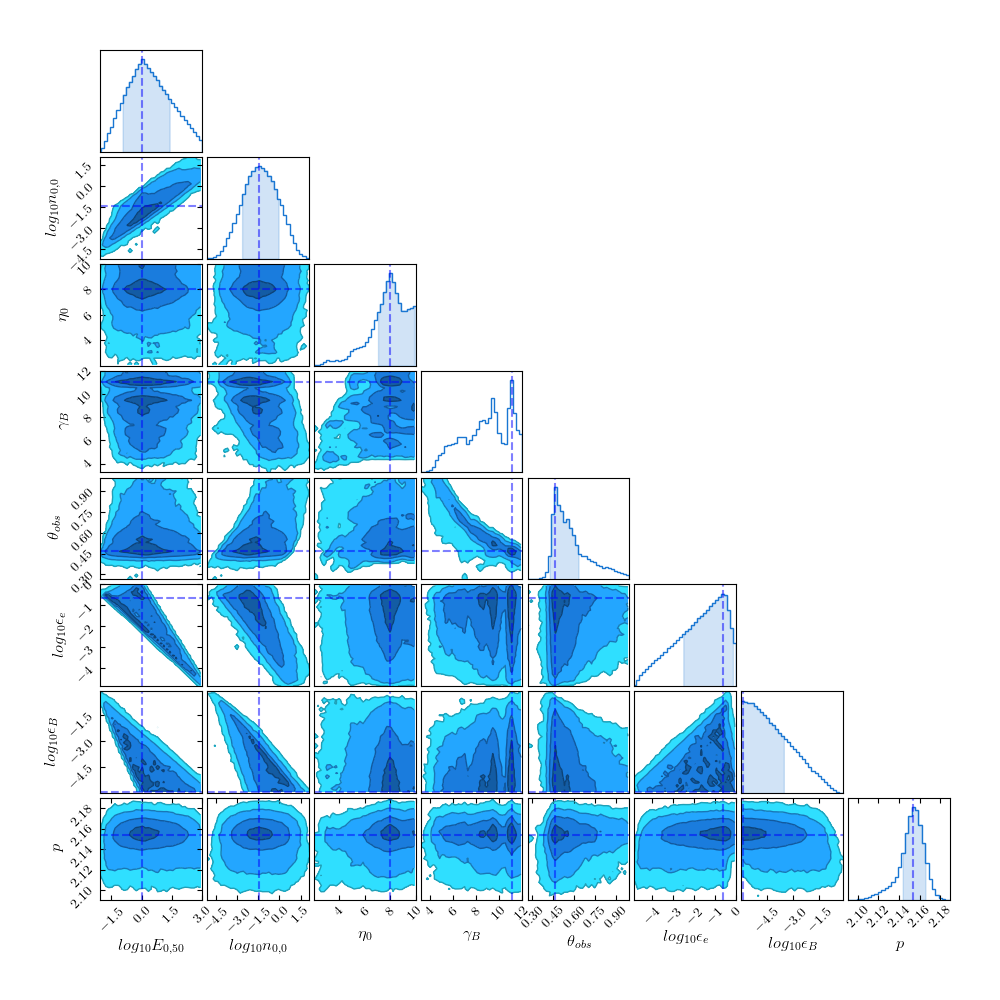}
\caption{The one dimensional (diagonal) and two dimensional (off-diagonal) projections of the posterior distributions. The dashed lines and shaded regions mark the medians and the symmetric $68\%$ uncertainties respectively.}
\end{figure*}

The fitting results are shown in Figure \ref{fig: distribution_off}. The corner plot shows the one dimensional marginalized posterior distribution (diagonal) for each parameter and the two dimensional posterior surface (off-diagonal) for  each pair of parameters. The dashed lines indicate the medians of the marginalized distributions and the symmetric $68\%$ uncertainties are shaded blue. The constraints of the parameters are summarized in Table \ref{tab:model_params_off}. We find  asymptotic Lorentz factor $\eta_0 \sim 8$ and boost Lorentz factor $\gamma_B \sim 11$ corresponding to characteristic jet opening angle  $\theta_0 \sim 1/\gamma_B \sim 0.09$ rad $\sim 5$ deg and the characteristic Lorentz factor  $\Gamma \sim  2\eta_0\gamma_B \sim 175$. This small jet opening angle and a large Lorentz factor indicates a collimated ultra-relativistic outflow. We find observation angle  $\theta_{obs} = 0.47^{+0.17}_{-0.05}$ rad $= 27^{+9}_{-3}$ deg, which indicates the center of the outflow is not oriented directly along our line of sight.  The analysis of the LIGO/Virgo gravitational wave data, independent of the radio/X-ray data, indicates that the inclination angle of GW170817 is $32^{+10}_{-13}$ deg \citep{Finstad2018}, which is consistent with our results. 

\begin{table}[hbt]
    \centering
    \caption{Median Parameter Values and their Bounds. }
    \label{tab:model_params_off}
    \begin{tabular}{lc}
        \hline
		Parameter & Median \\ 
		\hline
		$log_{10}E_{0,50}$ & $0.04^{+1.36}_{-0.98}$  \\ 
		$log_{10}n_{0,0}$ & $-1.4^{+1.4}_{-1.2}$  \\ 
		$\eta_0$ & $8.00^{+1.88}_{-0.94}$  \\ 
		$\gamma_B$ & $11.06$  \\ 
		$\theta_{obs}$ & $0.47^{+0.17}_{-0.05}$  \\ 
		$log_{10}\epsilon_e$ & $-0.65^{+0.49}_{-1.87}$  \\ 
		$log_{10}\epsilon_B$ & $-5.9^{+2.4}_{-0.0}$  \\ 
		$p$ & $2.154^{+0.012}_{-0.010}$  \\ 
		\hline
    \end{tabular}
\end{table}

Figure \ref{fig: distribution_off}, shows a clear correlation between $\gamma_B$ and $\theta_{obs}$ from the posterior surface. This is due to the peak time following the relation, e.g. \citep{Troja2017}:
\begin{equation}\label{eq: relation}
t_{peak} \propto \left(\frac{E_{0,50}}{n_{0,0}}\right)^{1/3} (\theta_{obs} - \theta_{0})^{2.5}.
\end{equation}
As we lower  $\gamma_B$, the corresponding  $\theta_0$ becomes larger. To keep the peak time around $150$ days, the relation requires a larger  $\theta_{obs}$. The marginalized distribution for $\gamma_B$ exhibits a bimodal behavior, which indicates that $\gamma_B \sim 9.5$ can also fit the observational data well. However, this value also corresponds to a relativistic structured jet, since  a quasi-spherical outflow is typically characterized by a much lower  $\gamma_B \sim 1\text{-}2$. The spectral index is tightly constrained $2.154^{+0.012}_{-0.010}$, which is close to the value found be \citet{Margutti2018}. 

The distributions for $E_{0,50}$, $n_{0,0}$, $\epsilon_e$ and $\epsilon_B$ are quite broad, covering several orders of magnitude. This is due to the degeneracies between these parameters, as we can see strong correlations between them from the posterior surfaces. The degeneracies can also be explained though the scaling relations in Equation \ref{eq: scale}. We can always vary the values for $E_{0,50}$, $n_{0,0}$, $\epsilon_e$ and $\epsilon_B$ to get the same values for $t_{obs}$,  $F_{peak}$,  $\nu_c$ and  $\nu_m$ and thus the same spectrum.  We find that high $E_{0,50}$ and high $n_{0,0}$ are often associated with very low $\epsilon_B$. This is related  to the constraint on the cooling frequency $\nu_c$. Spectral analysis shows that $\nu_c$ is above the X-ray band for GRB170817A \citep{Alexander2018, Dobie2018}. As we can see in Equation \ref{eq: scale},  $E_{0,50}$, $n_{0,0}$ and $\epsilon_B$ can be varied to satisfy the observed constraint on  $\nu_c$ . 

Even though our MCMC fitting cannot independently constrain $E_{0,50}$, $n_{0,0}$, $\epsilon_e$ and $\epsilon_B$ due to degeneracies, other parameters, in particular  $\eta_0$, $\gamma_B$ and $\theta_{obs}$, are robustly constrained. One advantage of MCMC fitting is that degenerate parameters can be marginalized out and their uncertainties can be incorporated into the marginalized distributions for the parameters of interest.

\begin{figure}[hbt]
\label{fig: bestoff}
\centering
\includegraphics[width=\linewidth]{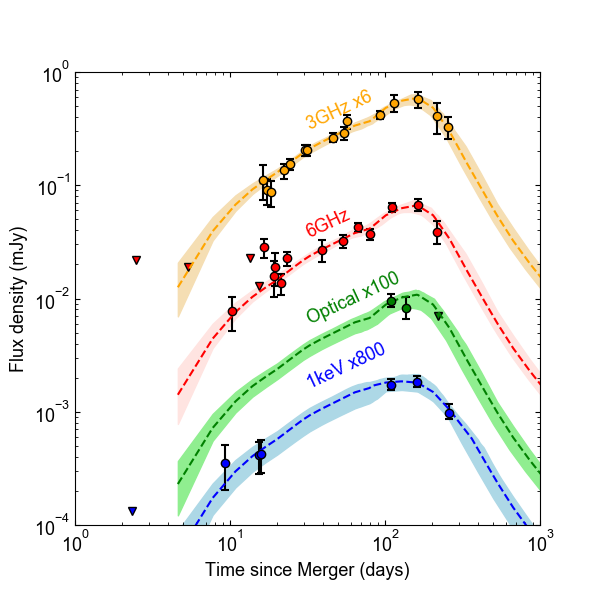}
\caption{The best-fitting light curves (dashed lines) are shown with the observational data up to $\sim260$ days, taken from \cite{Alexander2018,Margutti2018}. The triangles indicate the upper limits and the filled circles with error bars are the flux detections. The shaded regions indicate the range of light curves corresponding to the top $1\%$ of the MCMC samples. The reduced $\chi^2$ of the best-fitting light curves is $0.89$ (24.9/28).}
\end{figure}

We show the best-fitting light curves in Figure \ref{fig: bestoff}. The shaded regions demonstrate the light curves calculated from the models in the top $1\%$ of the MCMC samples. The reduced $\chi^2$ of the best-fitting light curves is $0.89\ (24.9/28)$. Between the mildly relativistic quasi-spherical outflow characterized by low $\eta_0$ and low $\gamma_B$ and the relativistic structured jet characterized by high $\eta_0$ and high $\gamma_B$, the MCMC analysis  favors the relativistic structured jet viewed from a significant off-axis angle. We also find the constraints for $\eta_0$, $\gamma_B$ and $\theta_{obs}$ are robust to variations of the degenerate parameters, such as $n_{0,0}$, as described in the following section.

\subsection{Fixed Circumburst Density}
For the full parameter space,  MCMC analysis cannot independently constrain $E_{0,50}$, $n_{0,0}$, $\epsilon_e$ and $\epsilon_B$,  due to  degeneracies mentioned previously. The marginalized distributions for these parameters are broad. For example, the distribution of $n_{0,0}$ ranges from $\sim 10^{-4}$ to $\sim 10^{1}$. Good fits to the observational data are possible  not only for low $n_{0,0}$, but also for fairly high $n_{0,0}$.  For the existing studies of GW170817, most successful models share a preference for low circumburst densities $n_{0,0} \sim 10^{-5} \text{-} 10^{-3}$ \citep{lazzati2017,Margutti2018,mooley2018mildly}. The best-fitting parameters for these models are usually determined by arbitrarily fixing some parameters and manually varying some other parameters.  $\epsilon_B$ is usually fixed at $\sim 10^{-3} \text{-} 10^{-1}$. As we have discussed in Section \ref{subsec: full}, a high $\epsilon_B$ may  lead to a low $n_{0,0}$ due to the cooling frequency constraint. Without using MCMC analysis, these models only explore a small part of the whole parameter space. Even though a certain set of parameters can provide a good fit, these models may neglect other sets of parameters, which can also give a comparably good fit, or even a better fit. 

\cite{Resmi2018} performed an MCMC analysis using a semi-analytic Gaussian jet model and found the following parameters: $E_{iso,c} = 10^{51.76}$ erg, $n_0 = 10^{-2.68}$ $\text{cm}^{-3}$, $\Gamma_c = 215$, $\theta_c = 6.9$ deg, $\theta_{obs} = 27$ deg, $\epsilon_e = 10^{-0.66}$, $\epsilon_B = 10^{-4.37}$ and $p = 2.17$. Their best-fitting parameters are consistent with our results, which also supports a relativistic structure jet viewed from a significant off-axis angle. By exploring the full parameter space with MCMC analysis, \citet{Resmi2018} also found a relatively high $n_0$ and a relatively low $\epsilon_B$. Thus, the circumburst density is not strictly required to be very low. 

\begin{figure*}[hbt]
\label{fig: distribution_Rho}
\centering
\includegraphics[width=\linewidth]{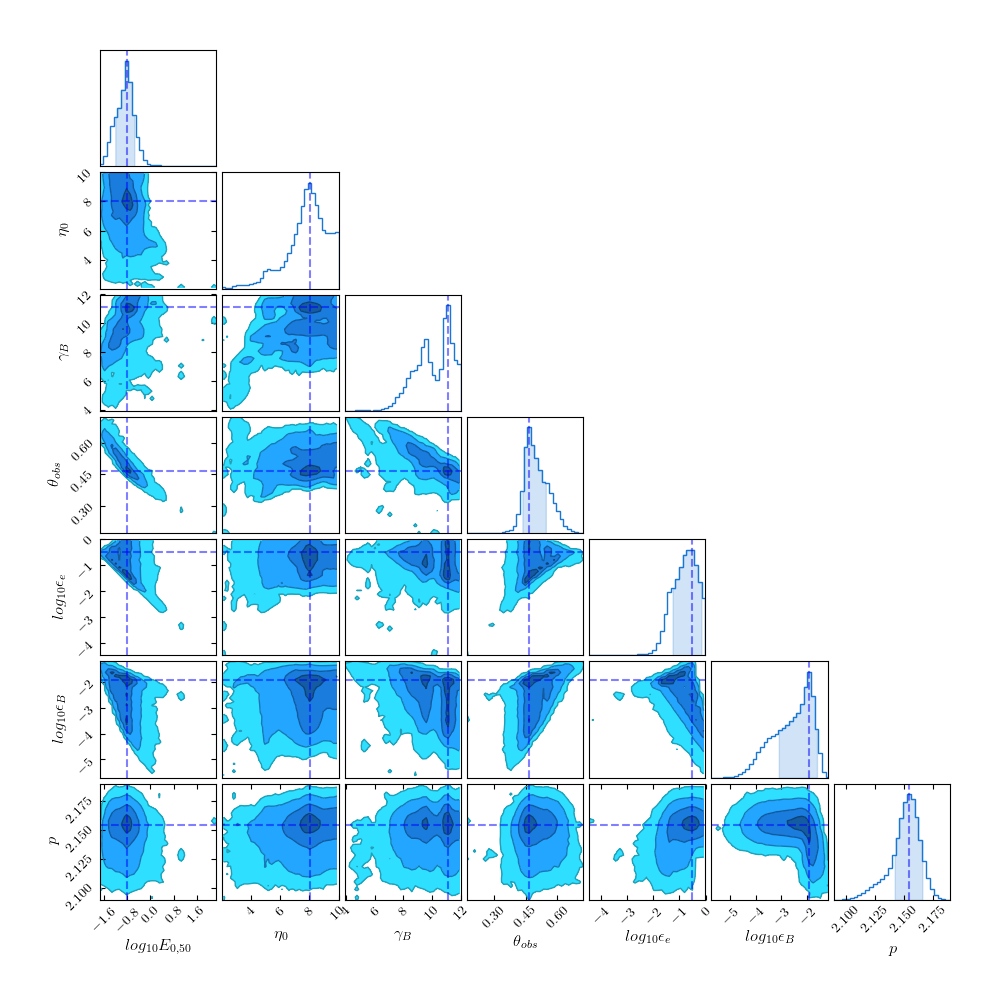}
\caption{The one dimensional (diagonal) and two dimensional (off-diagonal) projections of the posterior distributions. The dashed lines and shaded regions mark the medians and the symmetric $68\%$ uncertainties respectively. The parameter space has 7 dimensions with a fixed circumburst density $n_{0,0}=10^{-3}$.}
\end{figure*} 

\begin{table}[hbt]
    \centering
    \caption{Median Parameter Values and their Bounds.}
    \label{tab:model_params_Rho}
    \begin{tabular}{lc}
        \hline
		Parameter & Median \\ 
		\hline
		$log_{10}E_{0,50}$ & $-0.81^{+0.26}_{-0.39}$  \\ 
		$\eta_0$ & $8.00$  \\ 
		$\gamma_B$ & $11.11$  \\ 
		$\theta_{obs}$ & $0.47^{+0.08}_{-0.03}$  \\ 
		$log_{10}\epsilon_e$ & $-0.51^{+0.35}_{-0.75}$ \\ 
		$log_{10}\epsilon_B$ & $-1.91^{+0.30}_{-1.18}$  \\ 
		$p$ & $2.154\pm 0.012$  \\ 
		\hline
    \end{tabular}
\end{table}

To compare with other successful models with low circumburst densities, we fix $n_{0,0} = 10^{-3}$ and run MCMC fitting in the remaining 7-dimensional parameter space. The best-fitting light curves are almost identical to the light curves generated from the 8-dimensional parameter space and have a comparable reduced $\chi^2 = 0.86\ (24.9/29)$. The marginalized  distributions and posterior surfaces are shown in Figure \ref{fig: distribution_Rho}. The constraints are summarized in Table \ref{tab:model_params_Rho}. We find the following results: $\eta_0 \sim 8$,  $\gamma_B \sim 11$,  $\theta_0$ $\sim 5$ deg,   $\Gamma \sim  175$ and  $\theta_{obs} = 27^{+9}_{-3}$ deg. The consistent fitting results confirm that the MCMC method places robust constraints on $\eta_0$, $\gamma_B$ and $\theta_{obs}$. We find $\eta_0$, $\gamma_B$ and $\theta_{obs}$ are slightly better constrained, when we remove the uncertainty in $n_{0,0}$.  Fixing $n_{0,0}$ also helps to break the degeneracies between $E_{0,50}$, $n_{0,0}$, $\epsilon_e$ and $\epsilon_B$. We find improved constraints on $E_{0} \sim 2\times 10^{49}$ erg, $\epsilon_e \sim 0.3$ and $\epsilon_B \sim 0.01$. According to Equation \ref{eq: Eiso}, the isotropic equivalent energy can be estimated as  $E_{iso} \sim 4 \times 10^{52}$ erg. By fixing the circumburst density to $n_{0,0} = 10^{-3}$, the corresponding magnetic energy fraction  is driven to a larger value $\epsilon_B \sim 0.01$, due to the cooling frequency constraint. In a recent work, \citet{Gill2018} demonstrated light curves from the Gaussian jet model and the power-law jet model both with a fixed circumburst density $n_{0,0} = 10^{-3}$. By varying model parameters,  they found both models have similar results:  $\theta_c \sim 5$ deg,  $\theta_{obs} \sim 27$ deg,  $E_{iso,c} \sim 10^{52}$ erg, $\epsilon_B \sim 0.2$ and $\epsilon_B \sim 10^{-3}$. Their best-fitting parameters  are consistent with our results.

\section{Conclusions and Discussion}
\label{sec:discussion}
In this study, we present hydrodynamic simulations starting from a simple and physically motivated analytic outflow model, the two-parameter boosted fireball model, to investigate the outflow structure of GW170817. Parameterized by only two physical quantities, the asymptotic Lorentz factor $\eta_0$ and the boost Lorentz factor $\gamma_B$, the boosted fireball model serves as a generic model, which smoothly generates a family of outflow structures in the coasting phase. These include the two popular outflow structures, the mildly relativistic quasi-spherical outflow characterized by low $\eta_0$ and low $\gamma_B$ and the relativistic structured jet characterized by high $\eta_0$ and high $\gamma_B$ as well as intermediate outflow structures. 

We have run a total of $240$ relativistic hydrodynamic simulations starting with different values of $\eta_0$ and $\gamma_B$ and have calculated nearly $\sim 2,000,000$ spectra directly from these simulations using the standard synchrotron afterglow model. By making use of scaling relations in the radiation equations, we are able to generate tables for the characteristic spectral functions. With these tables, we generate synthetic light curves in milliseconds, which allows us to perform MCMC analysis in a reasonable amount of time. The light curve generator and the MCMC sampler provide a powerful tool to quickly investigate the outflow structure and the observation angle, not only for GW170817, but also for other neutron star merger events in the future. 

Using the tools we have developed, we fit the observational data of GW170817/GRB170817A up to $\sim 260$ days after the merger. Instead of manually varying a small set of parameters, our procedure fully explores the 8-dimensional parameter space and quickly finds the  best-fitting parameters and their distributions. We find asymptotic Lorentz factor $\eta_0 \sim 8$ and boost Lorentz factor $\gamma_B \sim 11$,  corresponding to characteristic jet opening angle  $\theta_0 $ $\sim 5$ deg and characteristic Lorentz factor  $\Gamma \sim 175$. The observation angle is found to be $\theta_{obs}$ $= 27^{+9}_{-3}$ deg. Our results thus favor the relativistic structured jet viewed from a significant off-axis angle rather than the mildly relativistic quasi-spherical outflow. 

We observe degeneracies between the explosion energy $E_{0,50}$, the circumburst density $n_{0,0}$, the electron energy fraction $\epsilon_e$ and the magnetic energy fraction $\epsilon_B$. MCMC analysis in the 8-dimensional parameter space cannot independently constrain these parameters. We find that a very low circumburst density $n_{0,0}$ is not strictly required to fit the observational data. The combination of high $n_{0,0}$ and low $\epsilon_B$ can also produce a comparably good fit. 
To compare with previous analyses, we fix  $n_{0,0} = 10^{-3}$ and perform MCMC fitting in the remaining 7-dimensional parameter space.  We find consistent results for  $\eta_0$, $\gamma_B$ and $\theta_{obs}$. Fixing $n_{0,0}$ also helps to break parameter degeneracies and achieve improved constraints on the values of  $E_{0} \sim 2\times 10^{49}$ erg, $\epsilon_e \sim 0.3$ and $\epsilon_B \sim 0.01$. 

In the near future, LIGO/Virgo will start O3, its third observing run, with an estimated BNS detection rate of up to $\sim1$ per month \citep{Abbott2017}. The light curve generator and the MCMC sampler presented here can be utilized to constrain the parameters of future afterglows from neutron star mergers including the outflow structure, circumburst density and observation angle. Observation angles constrained from future afterglow observations may aid in improving standard siren determinations of the Hubble constant \citep{1986Natur.323..310S,2010ApJ...725..496N} and can be utilized jointly in future gravitational wave analysis.

\acknowledgements
We are grateful to Xiaoyi Xie and Geoffrey Ryan for helpful discussions and comments. We thank Mulin Ding for maintaining the Ria  computing cluster. This research was supported in part by the NSF through grant AST-1715356.

\bibliographystyle{yahapj}
\bibliography{bibliography.bib}
\end{document}